# Station Reallocation and Rebalancing Strategy for Bike-Sharing Systems: A Case Study of Washington DC

Beigi, Pedram[1,3], Khoueiry, Michel [1], Rajabi, Mohammad Sadra[1] and Hamdar, Samer [2]
[1] PhD Student, Department of Civil and Environmental Engineering, The George Washington University, Washington DC, United States
[2] Associate Professor, Department of Civil and Environmental Engineering, The George Washington University, Washington DC, United States
[3] beigi@gwu.edu

**Abstract**: Bike-sharing is becoming increasingly popular as an urban traffic mode while increasing the affordability, flexibility, and reliability of interconnected public transportation systems (i.e., interconnected light rail, buses, micro-mobility, and ride-sharing modes of transportation). From the consumer's perspective, *1)* finding a bike station in convenient locations where demand usually occurs and *2)* the availability of bikes at rush hours with a lesser probability of encountering empty docks (for fixed-station bike-share systems) are two key concerns. Some stations are more likely to be empty or full, reflecting an imbalance in bike supply and demand. Accordingly, it is essential to understand a bike-share system's demand pattern to select the optimal locations and reallocate bikes to the right stations to increase the utilization rate and reduce the number of unserved customers (i.e., potential demand). The Capital Bikeshare in the Washington DC Metropolitan Area is one of the prominent bike-share systems in the USA - with more than 4,300 bikes available at 654 stations across seven jurisdictions. This study provides a systematic analysis of a bike-sharing system's Capital Bikeshare system usage pattern. Our study intends to create an optimization strategy formulated as a deterministic integer programming for reallocating bike stations daily and rebalancing the bike supply system. From an operational perspective, such a strategy will allow overnight preparations to answer the rush-hour morning demand and during special events in Washington D.C.

## 1. INTRODUCTION AND LITERATURE REVIEW

Bicycling has had a revival as a popular form of transportation and pleasure due to the rising priority that modern society places on health and sustainability (Erfani, et al. 2021) (García-Palomares, Gutiérrez and Latorre 2012). Capital bike share is one of the prominent bike-sharing systems in the United States, with more than 4,300 bikes available at 654 stations across seven jurisdictions.

When commuters face the usual last-mile difficulty, they confront when using just public transportation, they frequently opt for a motorized vehicle for the whole journey (Ahillen, Mateo-Babiano and Corcoran 2016). Bike-sharing systems (BSS) are intended to solve the last-mile problem by encouraging people to take public transit rather than driving (Shakerian, et al. 2022). This should result in more people taking public transportation. At the same time, it allows individuals to opt to cycle during rush hour if buses and subways are crowded, so it can solve the overcapacity problem.

It has been shown that one of the keys to the success of bike-sharing programs is the location of bike stations and their relation to trip demand (Lin and Yang 2011). (Todd, O'Brien and Cheshire 2021) described some of the variables that influence BSS usage, and explained that the Number of Stations, Station Density, Station Capacity, and Distance to Station are the most important ones. The placement of the station should be reliable and result in an acceptable total ride duration. Users prefer not to have to walk a long distance to drop off or pick up their bikes, but if there are too many stations close together, the cost of adding another station may outweigh the benefit of having that additional station (Beddis, Mitrovic and Sharma 2015). Arriving at a station with no available bikes or docks is a vital concern to BSS implementers. To better serve the users, policymakers should take note of this spatial mismatch trend and update the system setup accordingly (Taghavi, Khaleghparast and Eshghi 2021). The combination of station size and the defined maximum distance between stations and having a rebalancing strategy during day and night can reduce the likelihood of this issue (Mudiyanselage, et al. 2021).

(Bryant Jr 2013) employed heuristic spatial analytic approaches to solve a modified version of the set coverage problem in the city of Richmond, Virginia, as the maximum covering location problem to find the best places for a BSS. The algorithm was used to find bike stations within a certain distance to cover the most high-demand sites on a transit network. Bryant's model employed an origin-destination analysis (based on Dijkstra's algorithm) to determine a suitable walking distance between stations after identifying all current bus stops as prospective station locations. In the end, around 400 meters was decided.

(Croci and Rossi 2014) investigated the association between existing stations near points of interest and BSS usage, and they investigated the optimal re-location of present BSS stations in Milan using regression analysis results. (Ma, Liu and Erdoğan 2015) focused on the BSS's economic consequences based on station placements in Washington, D.C. (Beddis, Mitrovic and Sharma 2015) focus on maximizing the value of a BSS in the downtown Vancouver area by optimizing the placement of bike stations using data that provides pedestrian and bike traffic volumes and popular arrival and departure destinations.

(Rybarczyk and Wu 2010) and (Larsen, Patterson and El-Geneidy 2013) have utilized GIS-based multi-criteria analysis methods to assess bike infrastructure as well as estimate possible demand distribution. (García-Palomares, Gutiérrez and Latorre 2012) have shown the possibilities for location-allocation models integrated with GIS and applied to bike-sharing stations in the city center of Madrid.

(Kaviti, Venigalla and Lucas 2019) studied the effect of price change and travel behavior, which led to the realization that some stations are overfitted while others are under fitted. They also found that when stations are at full capacity, users feel discomfort since they cannot safely park the bike, while if it is empty, users have to walk an additional distance in order to find a useable bike. (Wergin and Buehler 2017) were able to deduce that different stations act as both origins and destinations at different times of day, which led to the realization that bike stations cannot be modeled as static entities but as dynamic objects that have different aspects depending on the time of the day.

Generally, modeling the location-allocation problem for a BSS is done in two ways: either (a) determining which locations were the most acceptable based on the distribution of criteria such as population and public transit density, or (b) creating a suitability map for locating the optimal locations utilizing optimization analyses such as asset or maximum covering problems. The first method aims to maximize the total number of people covered within a given radius by concentrating stations in areas with the highest potential demand. The second method allows station location to be maximized since it reduces the distance between supply and demand, resulting in a station distribution that is reasonably uniform over the whole area (García-Palomares, Gutiérrez and Latorre 2012).

This study aims to identify a set of bike-share station locations that will maximize the value of a public bike-share system and solve an optimization problem for rebalancing strategy. A deterministic integer-programming model is utilized to identify an appropriate selection of station locations for a BSS. The model takes into account a variety of parameters, including demand, station size, as well as extra values for station sites in places with improved accessibility and convenience.

## 2. METHODOLOGY

### 2.1. Model Considerations

The Capital Bikeshare system in the DC region, which has 334 stations in 179 census tracts, is the subject of this study. Potential station locations must be selected as part of the data collection process. At this point, we assume that all the intersections in Washington, DC, are potential stations; and after finding optimal stations, we walk through the target area and manually classify the locations that could fit a bike-share station near that intersection. The present bike-sharing system's ridership was utilized as the demand of each census tract to add value to station sites. It is preferable to assign a demand based on pedestrian and bike data, which is not available for Washington DC.

Considering other indications like stations near cultural places, tourist attractions, parks, and city parks were also deemed desirable to fulfill the user. As a result, cycling lanes, major tourism sites, popular locations, and subway stations were incorporated to better assess the value of each street for a potential BSS user. A station near bike lanes would be more valuable in terms of safety and convenience than a station on a typical street. (Buck and Buehler 2012) reveals a significant and positive relationship between bike lane supply near Capital Bikeshare stations and the number of trips originating from those stations.

The station density approach from previous successful studies such as New York City and Paris adheres to a density criterion of certain maximum and minimum distance between stations. (Nair et al., 2012) explain that Paris Vélib's docking stations are next to the Paris Metro stations and shows most frequently utilized docking stations are near transportation stops and services.

This model does not address station design and construction, bike-share equipment repair and maintenance, or determining bike rental pricing and payment methods. Our methodology focuses on identifying suitable station sites and their sizes. The tract demand distribution of the Capital Bikeshare system is presented in Figure 1(a), in which each dot represents bike where $D_j$ is defined as $\sum_{i \in j} \sum_k s_{pi} + s_{di}$ where $s_{pi}$ and $s_{di}$ represent the daily pick-up and drop off frequency for station $i$ in tract $j$ in day $k$. Capital Bikeshare system, net drop-off/pick-up in each tract are presented in Figure 1(b) and Figure 1(c) respectively.

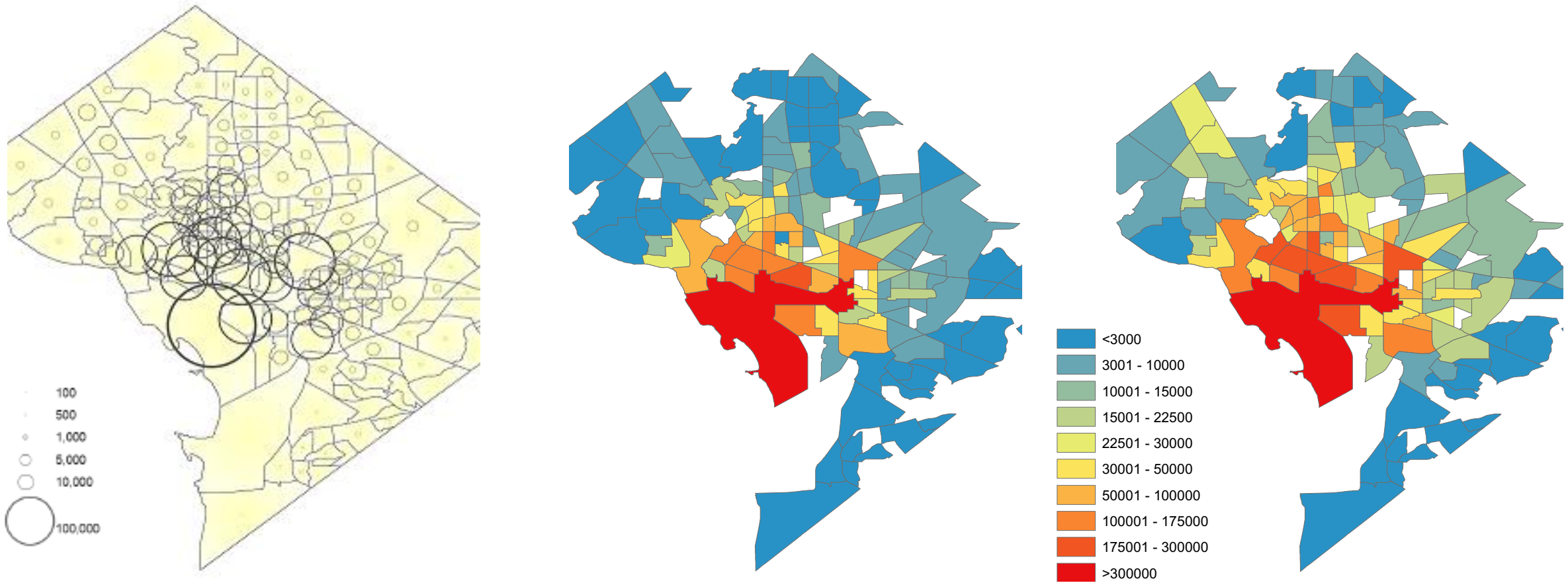

a) Demand Distribution b) Net-drop-up Distribution c) Net-pick-up Distribution

Figure 1: Bike demand distribution and balance distribution in District of Columbia

## 2.2. Station reallocation

The basic premise of our concept is that each potential station site has a corresponding value. The model sets a restricted number of stations to maximize the utility of the system, given specific limitations specifying where stations must or cannot be located.

Table 1: Parameters and Variables

| Type | Name | Explanation |
|---|---|---|
| Binary Variables | $x_i$ | Indicates whether a small-sized station is placed at location $i$ |
| Binary Variables | $y_i$ | Indicates whether a medium-sized station is placed at location $i$ |
| Binary Variables | $z_i$ | Indicates whether a large-sized station is placed at location $i$ |
| Sets | $P$ | Set of all potential locations |
| Sets | $F$ | Set of features that add value to locations |
| Variable | $N$ | Total number of stations to be placed |
| Variable | $M$ | Maximum number of medium stations |
| Variable | $L$ | Maximum number of large stations |

We assign location values by determining the demand and other relevant aspects of the target area. The features that we considered to add value to probable station sites are included in list $F$. Each feature $f \epsilon F$ has a value $v_f$ and is associated with a set of potential locations $P$. Demand distribution, bike lanes, protected bike lane, bike trail, bike shared lane, metro stations, and DC attractions are considered as features. Different size stations may be placed at each potential site in three sizes: small, medium, and large using three binary choice factors: $x_i$, $y_i$, *and* $z_i$. Furthermore, we verify that each potential site I has no more than one kind of station with $x_i + y_i + z_i \leq 1$ $for\ potential\ locations\ i$ and maximum number of stations has been considered as $\sum_{i \epsilon P} x_i + y_i + z_i < N$ where $N$ is the maximum number of stations placed. The most convenient solution for bike-share users would be a bike station installed at every potential location. However, this is not economically feasible; thus, the total number of stations that may be installed is restricted.

There is a list of neighbors for each potential location $j$, $Dj$ contains all possible station locations within $m$ meters of j, and $dist\left(x_i\,,\,x_j\right) \geq dm$ guarantees that if a station is placed at j, the nearest station to $j$ is at least $dm$ away. Similarly, if a station is established at position $j$, we do not want another station to be placed close next to it. $dist\left(x_i\,,\,x_j\right) \leq dl$ replicates the reality that clustering many stations in a limited region would not compound value for bike-share users.

In order to give the model an incentive to place larger stations near high-value features, we multiply the value of a potential station location by $\alpha$ if a small station is used, by $\beta$ if a medium station is used, and by $\gamma$ if a large station is used, where $\alpha\,,\beta,\gamma$ represents the scale of stations with $\sum_{i\epsilon P} z_i < L$ and $\sum_{i\epsilon P} y_i < M$ that ensure the number of medium and large stations are limited to $M$ and $L$. Therefore, our objective function is:

$$Maximize \sum_{i\epsilon P} v_f(\alpha x_i + \beta y_i + \gamma z_i)$$

### 2.3. Rebalancing strategy

This section aims to provide an optimization framework for the Capital Bikeshare Reorganizing System and to give the truck drivers an easy-to-follow schedule that outlines when and how to visit each station, as well as how many bikes they should pick up/deposit at each. We'll do so while attempting to reduce operational costs, which mostly comprise petrol costs, which vary depending on the distance traveled. To put it another way, we want to reduce the truck's effective distance traveled while also reducing the number of times it visits each station. We are assuming a truck passes 3 times per day (11 AM to reset after morning rush and prepare for evening rush, 7 PM to reset after evening rush and 3 AM to prepare for morning rush). On this basis, we divided our stations into 2 groups, one for Origins ($O$) and one for Destinations ($D$). The origin stations will be kept at 100% capacity in order to satisfy the biggest possible need, while the destination stations will be kept at 50% to accommodate for the small number of residents and miscellaneous trips while keeping empty docks so that people can safely deposit their bikes. The objective function is as follows while minimizing the total distance travelled:

$$Minimize \sum_{i,j} x_{ij} d_{ij}$$

Where $dij$ which represents the distance between origin station $i$ and destination station $j$. This distance is the Manhattan distance between station $i$ and $j$. Constraints $\sum_i x_{ij} \leq 1$ and $\sum_j x_{ji} \leq 1$ ensure that at most 1 trip departs from a station and at most 1 trip arrives to it, where $x_{ij}$ is a binary selection variable indicates a trips from station $i$ to station $j$. $\sum_i x_{ii} = 0$ ensures that even though the distance from a station to itself is always the shortest, it cannot be selected.

Constraint $b_i - o_i \leq p_i \leq b_i$ and $o_i - b_i \leq q_i \leq c_i - b_i$ show how many bikes should be pick up/drop at every station. Where $b_i$ represents the number of bikes at station $i$ before the truck arrives, $p_i$ is number of bikes picked up at station $i$, $q_i$ is number of bikes dropped at station $i$, parameter $o_i$ is the optimal number of bikes that should be available at station $i$ after the rebalancing process and $c_i$ is the capacity of station $i$. If we need to pick up, $t_j = t_i - q_i + p_i$ and $t_i \leq B$ show the number must be between the number of bikes in the station and the difference between this number and the optimal number of bikes in the station. If we need to drop bikes, the number must be between the difference of the current number of bikes in the station and the optimal number of bikes in the station and the difference between the number of bikes in the station and the maximum capacity of bikes, so we do not increase this number. Where variable $B$ is the maximum number of bikes we can fit on the truck and variable $t_i$ is the number of bikes on the truck before arriving to station $i$.

This specification can vary depending on the time of day, and we can get this feature by analyzing the data published on the Capital Bikeshare Website and check whether there is more departures or arrivals from the station in the desired timeframe. We can see the different number of trips departing and arriving on average in the year 2021. Figure 2 (a) shows the average number of trips during day. Comparing these values station by station would lead to determining the status of the station. Figure 2 (b) map shows the station type at a specific instant of time, with red signifying supply stations and blue demand stations.

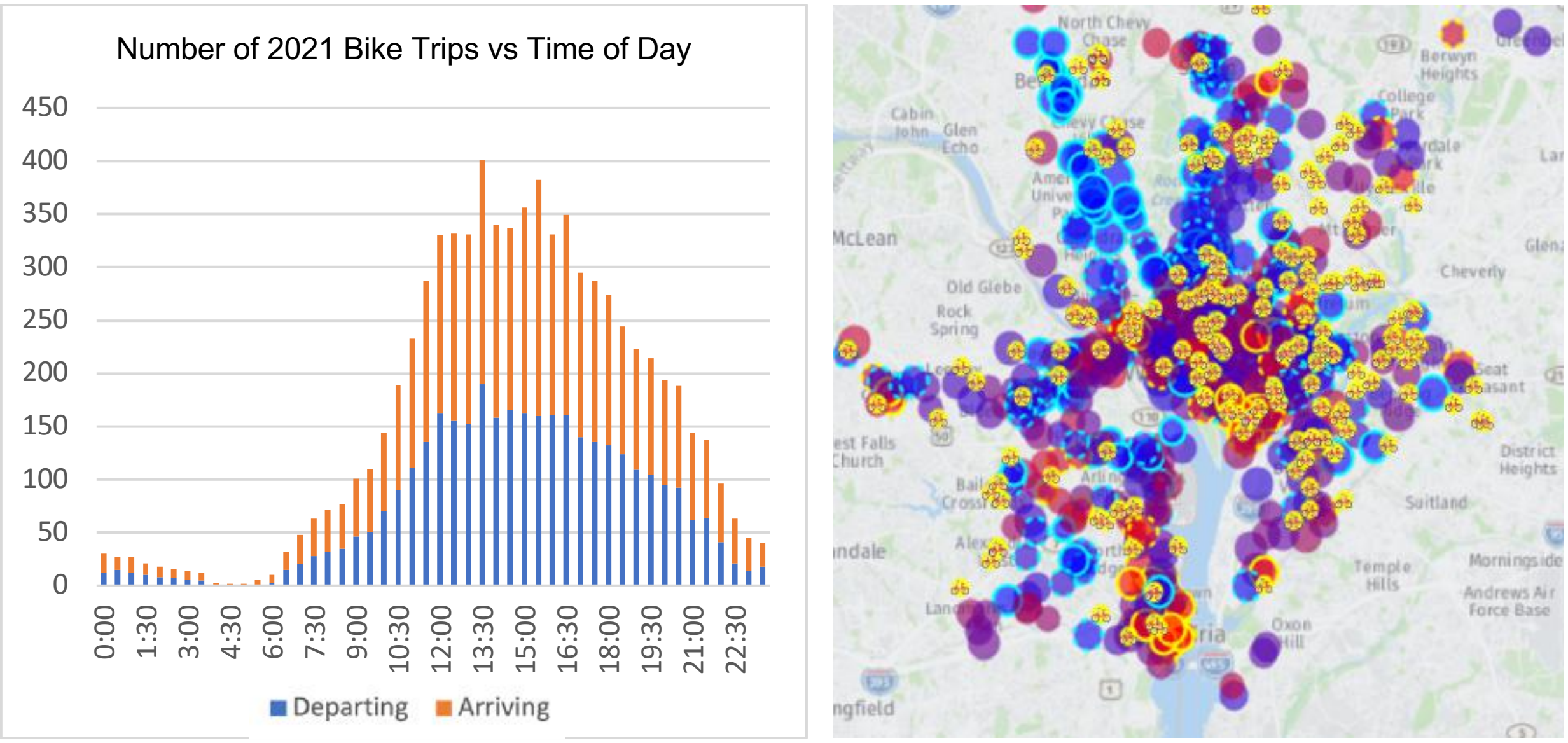

Figure 2: a) Average number of trips during day, b) Station type in DMV area

## 3. RESULTS AND DISCUSSION

We solved the model, considering our objective function and constraints and based on different number of $N$ and $m = 50$, $l = 50$, $dm = 300$, $dl = 1500$. Figure 3 illustrates the solution obtained. Decent set of bike station placements in the research region discovered which covered the study area, and there are no isolated clusters of stations that are difficult to reach or depart from, and the city's most popular and valuable areas are easily accessible. Varying the total number of stations to be placed is an interesting problem by itself. The more stations we are allowed to place, the higher our objective function value will be. However, given that certain locations are more valuable than others, we expect that there will be diminishing returns. That is, if $N$ is small, then it is likely that there is another highly valued location that can still be used, $N + 1$ will result in a significant rise in the objective function value. If $N$ is big, however, there are unlikely to be any more highly rated places to choose from, and so placing $N + 1$ stations will only result in a slightly higher objective function value.

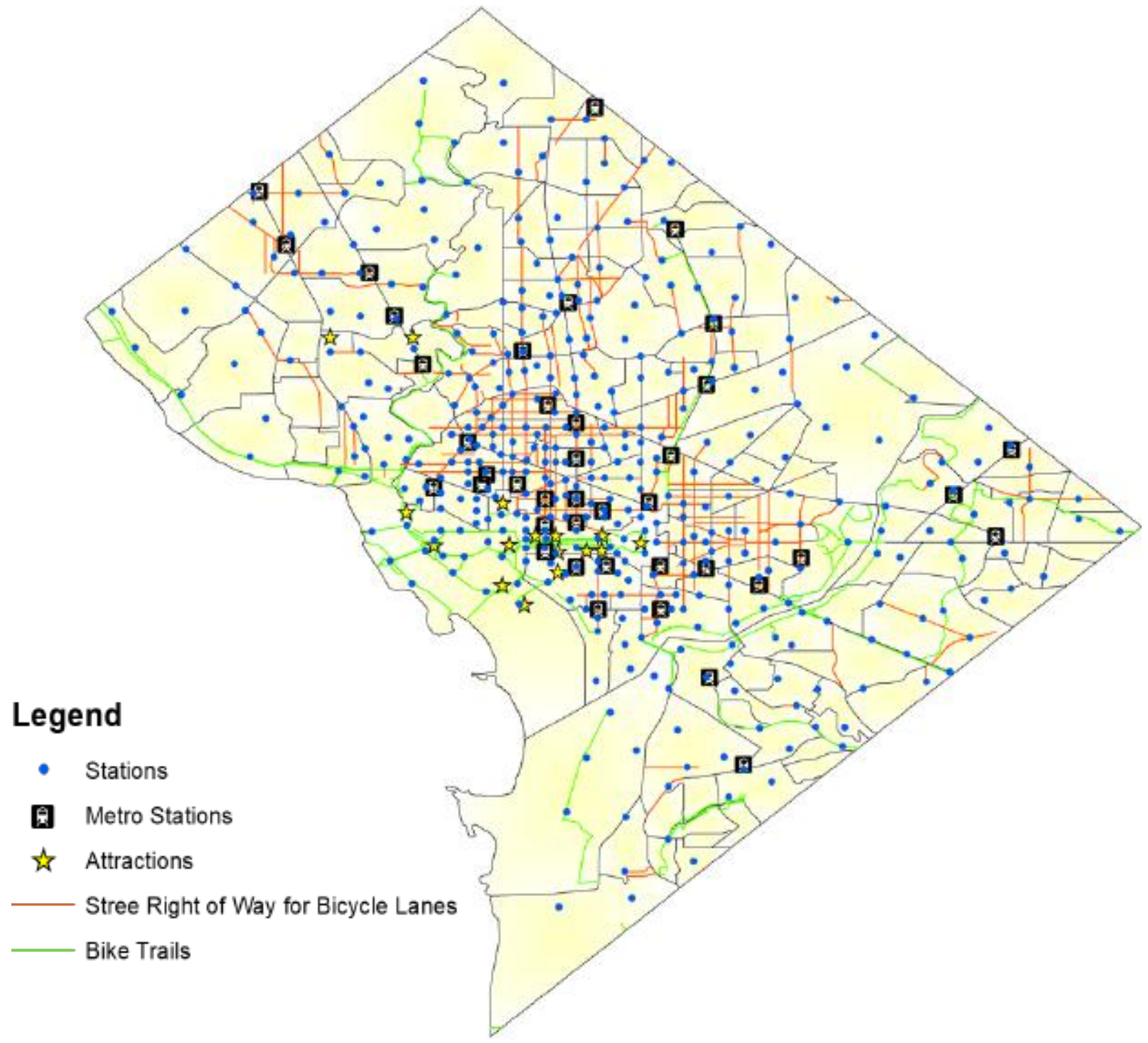

Figure 3: Station allocation result

We solve our model for $N$ values that steadily increase. Figure 4 (b) shows a plot of the objective function value vs number of stations, it indicates a falling rate of rise that virtually flattens out completely, as predicted. Because of the as-needed region restrictions, low values of $N$ are either infeasible or result in station placements that do not add much value to the objective function, the above graph does not consider small numbers of $N$. That is, before attempting to maximize the objective function, the model must first fulfill the constraints. This study supports the concept that $N$ = 400 is a suitable choice if adequate resources are available.

We were also able to solve the rebalancing model. Figure 4 (a) depicts the simple output that helps the driver visualize his tasks and itinerary for Foggy Bottom region, which contains 22 stations. As can be seen, our algorithms aid the driver in determining the number of bikes to be picked up (positive) or dropped off (negative), with a 0 value for the station indicating that no rebalancing is required with showing order of stations to be reach. The optimal method would be to divide the DC region into multiple zones based on the number of trucks and apply the technique to each zone separately.

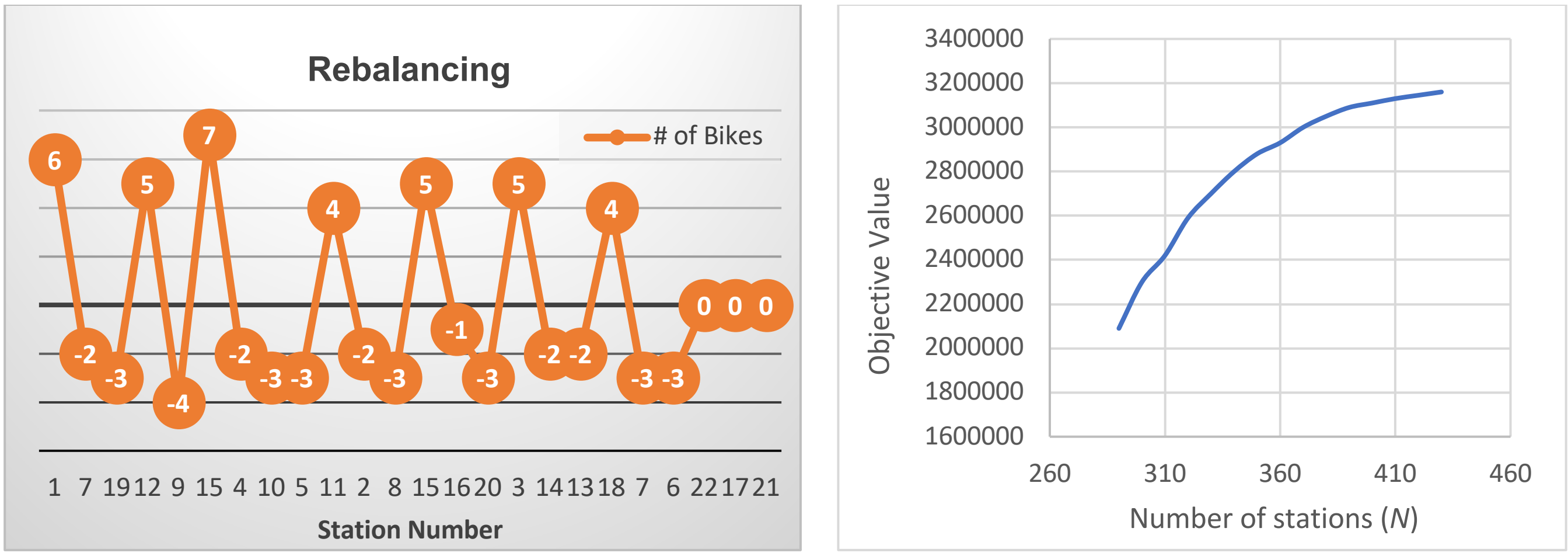

Figure 4: a) Visualization for rebalancing strategy, b) Increase rate of objective value by increasing of $N$

## 4. CONCLUSION

This study provided an optimization solution for a centralized bike-relocation system for Capital Bikeshare and developed a comprehensive bike station network optimization approach by selecting bike station locations with high value defining with features which led to improving the performance of the BSS system. The models presented can be readily replicated to other data sources or regions, and it holds the potential to provide implications for transportation planning and policy making.

The placement of bike stations relationship with demand and the public transportation system, are all important factors in the success of a BSS. Since all other infrastructures are linked to bicycles, new facilities are often developed for recreational cycling or keeping bicycles "out of the way" of motorized traffic in the absence of such a system for situating new facilities (García-Palomares, Gutiérrez and Latorre 2012). However, to best serve, methods should be developed to objectively determine how to optimally place a BSS (Larsen, Patterson and El-Geneidy 2013). Measuring the value of the objective function has made it possible to confirm that an increment in the number of stations logically leads to an increase in both the demand covered and the accessibility of stations to potential destinations but with diminishing returns. As (De Chardon, Caruso and Thomas 2016) note, a large number of stations may result in an unreasonable rise in the system's cost without bringing about any significant improvements.

The present study has some limitations. The demand distribution will benefit from pedestrian and bike data and empty/full station patterns analysis could benefit from a complete data set covering a more extended period. Due to data unavailability, the optimization analysis does not consider the dynamic demand in real-time but rather uses past trip records to approximate demand. In addition, the optimization analysis would become more practical if the costs of adding a station and shifting a dock between stations are available. A possible improvement to the model would be to include the BSS revenues and expenditures, which could help to optimize the results further. This issue would be addressed by requesting more data from the Capital Bikeshare System in future studies. Also, the connection of the installed stations is a potential improvement to our approach. While we ensure that each station is at most 300 meters from another, that instance does not rule out the possibility of isolated clusters of stations apart from the main set. Another limitation is the presence of relatively isolated stations, particularly in the maximize-coverage model. However, these stations can easily be identified and eliminated after the accessibility analysis (García-Palomares, Gutiérrez and Latorre 2012).